\documentstyle[pra,aps,epsfig]{revtex}
\def  \dag{{\dagger}}
\begin{document}
\flushbottom
\draft
\title{Nonlinear optics of matter
waves \footnote{Dedicated to Peter Franken on his 70th Birthday}}
\author{E. V. Goldstein, M. G. Moore, O. Zobay, and P. Meystre}
\address{Optical Sciences Center and Department of Physics\\
University of Arizona, Tucson, Arizona 85721
\\ \medskip}\author{\small\parbox{14.2cm}{\small \hspace*{3mm}
We give a brief overview of the way atomic physics is now developing in
a way reminiscent of the optics revolution of the 1960's. Thanks in
particular to recent developments in atomic trapping and cooling, the new
field of atom optics is rapidly leading to exciting new developments
such as nonlinear atom optics and quantum atom optics. We illustrate these
developments with examples out of our own research.
\\[3pt]PACS numbers: 03.75.-b,42.50.Ct,42.50.Vk,42.65.Hw}}
\maketitle

\section{Introduction}

Atom optics has witnessed considerable progress in the last few years.
A number of optical elements, including atomic mirrors \cite{BalLet89}
and gratings \cite{MosGouAtlPri83}, have
been demonstrated, and several types of atom interferometers have been
built \cite{AppPhys92,AdaSigMly94,AdaCarMly95}.
In addition, advances in laser cooling and trapping of atoms
\cite{ChuBjoAsh86,PhiGou87,LetWatWes88,ChuWie89,DalCoh89,CohPhi90}
have led to spectacular developments, including the observation of quantized
atomic motion in optical lattices\cite{JesGerLet92,GryLouVer93},
and, in the last few years, the
demonstration of Bose-Einstein condensation
\cite{AndEnsMat95,MewAndDru96,BraSacHul97}
and the realization of a
primitive atom laser \cite{MewAndKur97,AndTowMie97}.
Atom optics using condensates as sources has now
become reality, as illustrated, e.g., by experiments involving the interference
of two condensates \cite{AndTowMie97}, the Kapitza-Dirac
diffraction of a condensate off a
standing-wave grating \cite{OvcKozDen98},
or proposed experiments in matter-wave phase
conjugation \cite{GolMey983} and matter-wave amplification
\cite{MooMey983,LawBig98}.

The early experiments in atom optics considered low
density samples, where atom-atom interactions are negligible and the atoms
in the beam behave independently. We call this regime {\em linear atom
optics.} When atom-atom interactions become important, the dynamics of
a given atom in the sample becomes dependent on the presence of other
atoms, and one reaches the regime of {\em nonlinear atom optics}
\cite{LenMeyWri93,ZhaWalSan94,LenMeyWri94}, the
matter waves analog of nonlinear optics, which was pioneered by Peter
Franken in 1962 \cite{FraHilPet61}.

In addition to making a distinction between linear and nonlinear atom
optics, it is important to also separate the {\em ray optics} from the
{\em wave optics} regime: In the first case, it is sufficient to
describe the atoms as point particles following classical trajectories
influenced by light fields. Much of laser cooling belongs to that category.
In contrast the wave atom optics regime, which is normally --- but not
exclusively --- associated with ultracold atomic samples, requires a
proper quantum mechanical description of the center-of-mass motion of
the atoms, including the effects of matter-wave diffraction.

Finally, a last distinction of importance is between what might be called
the {\em quantum} and {\em classical} regimes of atom optics, again in 
analogy with the electromagnetic case: the quantum regime of both
atom optics and optics is that regime where the effects of quantum
statistics play a significant role.  Similarly to optics, where both the
classical and
the quantum field are governed by Maxwell's equations, both the ``classical''
and the second-quantized Schr\"odinger fields are governed by the
Schr\"odinger equation.

The quantum regime of matter-wave optics is conveniently
described within the framework of second quantization of the matter wave
field, which is described ``classically'' by the wave function
$\psi({\bf r}, t)$, so that $\psi({\bf r}, t) \rightarrow
{\hat \Psi}({\bf r}, t)$. This is analogous to the optical case,
where the electromagnetic field is quantized, $E({\bf r}, t) \rightarrow
{\hat E}({\bf r}, t)$. It should be remarked, however, that
in contrast to optics,
where the quantization of the field introduces new physics, such is
not the case when second-quantizing the Schr\"odinger field at the low
energies considered here. In this regime, the total number of particles
is of course conserved, since effects such as pair creation are ignored.
Hence, for our purpose the second quantization procedure is merely a
convenient book-keeping mechanism which automatically accounts for particles
being added and removed from a given state with the proper quantum
statistics. Note however that we intentionally distinguish the quantum from
the nonlinear regime of atom optics, since reaching the first one requires
that a condition on phase-space density be met, while the second one is related
to density only.

The goal of this paper is to illustrate the recent developments in quantum
nonlinear atom optics with the help of three specific examples. We first
recall in Sec.~II how collisions are the main source of
nonlinearities in atom optics, showing the parallels
between this and the situation in nonlinear optics. Section III presents
a matter-wave four-wave mixing process, the generation of a phase
conjugate (time-reversed) matter wave in multicomponent condensates.
Section IV briefly reviews a proposed scheme of an
atom laser, and Sec.~V discusses a parametric process taking place
in matter waves, the so-called collective atomic recoil laser. Finally,
Sec.~VI is a summary and outlook.

\section{The role of collisions}

The analogy between nonlinear atom optics and nonlinear optics
is quite profound: In conventional optics, effective nonlinear
equations for the optical fields result from the elimination of the
medium dynamics, while in atom optics, nonlinear matter-wave dynamics
results from collisions, which are in turn an effective manybody
interaction resulting from the (partial) elimination of the
electromagnetic field. Indeed, at the most fundamental level of quantum
electrodynamics there is no such thing as two-body interactions between
atoms; rather, the potentials that describe them are the result of a series of
approximations whose validity depends on the precise situation at hand.

One simple way to illustrate how this works is to consider the near-resonant
dipole-dipole interaction inside a cavity.
Consider two two-level atoms of Bohr transition frequency
$\omega_a$, and located at positions ${\bf x}_1$ and ${\bf x}_2$,
with $|{\bf x}_1-{\bf x}_2|=x$ along the axis of a ring cavity of
length $L$. This system is described by the Hamiltonian
\begin{equation}
H=H_a+H_{af}+H_f,
\label{ham}
\end{equation}
where the atomic Hamiltonian $H_a$ is the sum of the
individual atomic Hamiltonians
\begin{equation}
H_a=\sum_{i=1,2} H_a^{(i)}=\sum_{i=1,2}\left ( \frac{{\bf p}_i^2}{2M} +
\hbar \omega_a \sigma_+^{(i)} \sigma_-^{(i)} \right ),
\label{hama}
\end{equation}
${\bf p}_i$ are the momenta of the atoms of mass $M$,
with $[{\bf x}_i, {\bf p}_j] = i\hbar
\delta_{ij}$,  and $\sigma_+^{(i)}$ and $\sigma_-^{(i)}$ are the atomic raising
and lowering pseudo-spin operators for the $i$-th atom.
The field Hamiltonian is
\begin{equation}
H_f=\sum_{\mu}\hbar\omega_\mu a^\dagger_\mu a_\mu,
\label{hamf}
\end{equation}
where the composite index $\mu=\{{\bf k}_n,\ell\}$ labels both the wave number
${\bf k}_n$, with $k_n=\omega_n/c$, and polarization
$\ell$ of the cavity mode of frequency $\omega_n$. The creation
and annihilation operators $a^\dagger_\mu$ and $a_\mu$ satisfy the
usual Bose commutation relation $[a_\mu,a^\dagger_{\mu'}]=\delta_{\mu \mu'}$.
Finally, the atom-field interaction Hamiltonian $H_{af}$ is the sum of
the individual atom-field electric dipole interaction Hamiltonians
\begin{equation}
H_{af}^{(i)}=-{\bf d}_i\cdot {\bf E}({\bf x}_i),
\label{hamaf}
\end{equation}
where
\begin{equation}
{\bf d}_i=\hat{\mbox{\boldmath {$\epsilon$}}}d(\sigma_+^{(i)}+
\sigma_-^{(i)})
\label{dip}
\end{equation}
is the atomic dipole operator, aligned along
$\hat{\mbox{\boldmath {$\epsilon$}}}$ and of magnitude $d$, and
\begin{equation}
{\bf E}({\bf r})=i\sum_\mu {\cal E}_n[a_\mu e^{i{\bf k}_n\cdot{\bf r}}
\hat{\mbox{\boldmath {$\epsilon$}}}_\ell-H.c.].
\label{field}
\end{equation}
Thereby,
${\cal E}_n$ is the electric field per photon in mode $\mu$, whose explicit
form depends upon the choice of quantization scheme \cite{Coh92,MeySar91}.
For a running wave quantization scheme and periodic boundary conditions
appropriate for a ring cavity we have in one dimension
${\cal E}_n= [\hbar\omega_n/(2\epsilon_0L)]^{1/2}$,
where $\omega_n=2\pi nc/L$. In the rotating-wave approximation,
the atom-field Hamiltonian (\ref{hamaf}) reduces then to
\begin{equation}
H_{af}^{(i)}=-i\hbar\sum_\mu\left [ g_{\mu i}(x_i) a_\mu \sigma_+^{(i)}
-g_{\mu i}^\star(x_i) a_\mu^\dagger \sigma_-^{(i)}\right ],
\label{rwa}
\end{equation}
where
\begin{equation}
g_{\mu i}(x_i) =
({\cal E}_n d/\hbar)(\hat{\mbox{\boldmath {$\epsilon$}}}_\ell
\cdot \hat{\mbox{\boldmath {$\epsilon$}}})\exp(i{\bf k}_n\cdot {\bf x}_i).
\end{equation}

We ignore in this section the kinetic energy part in 
the Hamiltonian (\ref{hama}),
and consider the situation where one of the atoms is initially excited,
the other is in its ground state, and the radiation field is in the
vacuum state,
\begin{equation}
|\psi(0)\rangle=|eg0\rangle.
\label{inst}
\end{equation}
In the rotating-wave approximation, the Hamiltonian (\ref{ham}) conserves
the number of excitations in the system, so that at time $t$ its state
vector can be expressed as
\begin{equation}
|\psi(t)\rangle=b_1(t)|eg0\rangle+b_2(t)|ge0\rangle+
\sum_\mu b_\mu(t)|gg1_\mu\rangle,
\label{state}
\end{equation}
with $b_1(0)=1$ and $b_2(0)=b_\mu(0)=0$. 

From the Schr\"odinger equation $i\hbar|\dot{\psi}(t)\rangle=H|\psi(t)\rangle$, 
the equations of motion for the various probability amplitudes involved are
readily found to be
\begin{eqnarray}
\dot{b}_i(t)&=&\sum_\mu g_{\mu i} b_\mu(t),
\nonumber \\
\dot{b}_\mu(t)&=&-i\Delta_n b_\mu(t)-g^\star_{\mu 1}{b}_1(t)
-g^\star_{\mu 2} {b}_2(t),
\label{system}
\end{eqnarray}
where $i=1,2$ and $\Delta_n\equiv \omega_n-\omega_a$ is the detuning of
mode $\mu$ from the atomic transition frequency.

The simplest way to solve the set of equations (\ref{system}) is to neglect
the effects of interatomic propagation. This is appropriate provided
that $x/c \ll \Gamma^{-1}$, where $\Gamma$ is the single-atom
free-space spontaneous decay rate and $x$ is an interatomic separation.
{\em Assuming that the atoms interact with a broadband vacuum and that 
the Born-Markov approximation holds}, one then finds
for the  probability amplitudes involving excited-state atoms \cite{MilKni74}
\begin{equation}
b_{1,2}(t)=\frac{1}{2}\left [ C_+(t) \pm C_-(t) \right ] ,
\label{coef}
\end{equation}
where
\begin{equation}
C_\pm(t)=\exp\left\{-\frac{1}{2}[\Omega_s\pm\Omega_{12}(x)]t\right\} .
\label{ct}
\end{equation}
Here we have introduced the {\em single-atom} and the {\em two-body} complex
frequencies $\Omega_s$ and $\Omega_{12}(x)$. The explicit form of
$\Omega_s$ is
\begin{equation}
\Omega_s=\Gamma+i\Delta_s,
\end{equation}
where
\begin{equation}
\Gamma=2\sum_\mu|g_{\mu i}|^2\delta(c| k|-\omega_a) =
\frac{L}{2\pi}\frac{d^2}{\hbar\epsilon_0L}\int_{-\infty}^\infty
dk c |k| \pi \delta(c|k|-\omega_a)
=\frac{d^2\omega_a}{\hbar\epsilon_0 c}
\label{gamma}
\end{equation}
is a one-dimensional version of the free-space spontaneous
emission rate, and
\begin{equation}
\Delta_s=-\Gamma \frac{P}{\pi}\int d\omega \frac{\omega}{\omega_a}
 \frac{1}{\omega-\omega_a}
\label{deltas}
\end{equation}
is the one-dimensional, two-level atom version of a Lamb shift.

More interesting in the present context is the two-body complex
frequency $\Omega_{12}(x)$ is defined as
\begin{equation}
\Omega_{12}(x)\equiv\Gamma_2(x)+iV_{dd}(x)/\hbar
\label{Omega}
\end{equation}
and its real part $\Gamma_2(x)$ accounts
for the modulation of the spontaneous decay rate of one of the atoms due to
the presence of a second atom at a distance $x$, while its imaginary part is
proportional to the dipole-dipole interaction potential $V_{dd}(x)$. For the
one-dimensional situation considered here, one finds explicitly
\begin{equation}
\Omega_{12}(x)=\frac{d^2}{2\pi\epsilon_0\hbar}\int_{-\infty}^{\infty}
dk c |k|e^{ikx} \left[\pi\delta(c|k|-\omega_a)-iP\left(\frac{1}{c|k|-\omega_a}
\right)\right],
\label{omt}
\end{equation}
so that
\begin{equation}
\Gamma_2(x)= \Gamma \int_0^\infty d \omega \frac{\omega}{\omega_a}
\delta(\omega - \omega_a)
= \Gamma \cos(k_ax)
\label{gamma2}
\end{equation}
and
\begin{equation}
V_{dd}(x)= -\hbar \Gamma \frac{P}{\pi}\int_0^\infty d\omega
\frac{\omega}{\omega_a}\frac{ \cos(\omega x/c)} {\omega - \omega_a}
= -\hbar\Gamma\sin(k_ax),
\label{vdd}
\end{equation}
where $k = \omega/c$ and $k_a=\omega_a/c=2\pi/\lambda_a$.
For the more familiar
three-dimensional case, the trigonometric functions appearing in these
expressions are replaced by combinations of Bessel functions, and
$V_{dd}(r)$ falls off as $1/r$ for large interatomic separations, instead of
being periodic \cite{Ste64,SmiBur91,LenMey93}.
Note also that as a result of the Born-Markov
approximation, these expressions neglect propagation and are independent
of time.

We see, then, that the two-body dipole-dipole interaction between atoms
results as advertised from the adiabatic elimination of the dynamics of the
electromagnetic field. This is analogous to the nonlinear optics
situation, except that the roles of the atoms and
the electromagnetic field are reversed.

In the following sections we discuss several examples of quantum
nonlinear atom optics which show how the familiar ideas of nonlinear optics
can be readily transposed to the new situation. We consider first four-wave
mixing in a multicomponent Bose-Einstein condensate, which can lead, e.g.,
to matter-wave phase conjugation. We then discuss in Sec.~IV a form of
matter-wave ``amplification'' reminiscent of laser action in optics, and in
Sec.~V a parametric process, the low-temperature version of the
collective atom recoil laser. Finally, Sec.~VI is a conclusion and outlook.

\section{Four-wave mixing: matter-wave phase conjugation}

A particularly close analogy can be established between the
dynamics of a spin-1 multicomponent condensate, as recently realized in
sodium experiments and the situation of degenerate four-wave mixing in optics.
Specifically, in the zero-temperature limit a $q$-component condensate can
be thought of as a $q$-mode system, whereby the various modes are coupled
by two-body (and possibly higher-order) collisions which result in the
exchange of particles between these modes.

In particular, for the case of a $^{23}$Na condensate in an optical dipole
trap  one can achieve situations where the $m=0$ state is macroscopically
populated while the $m=\pm 1$ states are weakly excited. One can then think
of the first state as a ``pump'' or "central" mode, while $m=\pm 1$ form
side modes which are coupled via the pump. This leads to familiar effects such
as degenerate four-wave mixing and matter-wave phase conjugation.

Consider then a condensate of $^{23}$Na atoms in their $F=1$ hyperfine ground
state, with three internal atomic states $|F=1,m=-1\rangle$,
$|F=1,m=0\rangle$ and $|F=1,m=1\rangle$ of degenerate energies in the absence
of magnetic fields. It is described by the three-component vector
Schr\"odinger field
\begin{equation}
\bbox{\Psi}({\bf r},t)
=\{\Psi_{-1}({\bf r},t),\Psi_{0}({\bf r},t),\Psi_{1}({\bf r},t)\}
\end{equation}
which satisfies the bosonic commutation relations
\begin{equation}
[\Psi_i({\bf r},t), \Psi_j^\dagger({\bf r}',t)]
=\delta_{ij}\delta({\bf r}-{\bf r}').
\end{equation}
Accounting for the possibility of two-body collisions, its dynamics
is described by the second-quantized Hamiltonian
\begin{equation}
{\cal H}=\int d {\bf r} \bbox{\Psi}^\dagger({\bf r},t)H_0 \bbox{\Psi}({\bf r},t)
+\int \{d {\bf r}\}\bbox{\Psi}^\dagger({\bf r}_1,t)
\bbox{\Psi}^\dagger({\bf r_2},t)V({\bf r}_1-{\bf r}_2)
\bbox{\Psi}({\bf r}_2,t)\bbox{\Psi}({\bf r}_1,t),
\label{ham2}
\end{equation}
where the single-particle Hamiltonian is
\begin{equation}
H_0={\bf p}^2/2M + U({\bf r})
\end{equation}
and the dipole trap potential $U({\bf r})$ does not depend on the hyperfine
magnetic state $m$.

The general form of the two-body interaction $V({\bf r}_1-{\bf r}_2)$
has been discussed in detail in Refs. \cite{Ho98,ZhaWal98}.
Labeling the hyperfine states of the combined system of hyperfine spin
${\bf F} = {\bf F}_1 + {\bf F}_2$ by $|f,m \rangle$ with $f = 0, 1,2$
and $m=-f,\ldots,f$, it can be shown that in the shapeless approximation
the two-body interaction is of the general form \cite{Ho98}
\begin{equation}
 V({\bf r}_1-{\bf r}_2)=\delta({\bf r}_1-{\bf r}_2)\sum_{f=0}^{2}
\hbar g_f {\cal P}_f,
\label{pot}
\end{equation}
where
\begin{equation}
g_f=4\pi\hbar a_f/M,
\end{equation}
${\cal P}_f\equiv \sum_m|f,m\rangle\langle f,m|$ is the projection operator
which projects the pair of atoms into a total hyperfine $f$ state
and $a_f$ is the $s$-wave scattering length for the channel of total
hyperfine spin $f$. For bosonic atoms only even $f$ states contribute, so
that
\begin{equation}
V({\bf r}_1-{\bf r}_2)=\hbar \delta({\bf r}_1-{\bf r}_2)(g_2{\cal P}_2+
g_0{\cal P}_0) = \frac{\hbar}{2}\delta({\bf r}_1-{\bf r}_2)
\left(c_0+c_2{\bf F}_1\cdot{\bf F}_2 \right).
\end{equation}
In this expression,
\begin{equation}
c_0 =2(g_0+2g_2)/3, \mbox{\hspace{5.mm}} 
c_2 = 2(g_2-g_0)/3.
\end{equation}
Substituting this form of $V({\bf r}_1-{\bf r}_2)$ into the second-quantized
Hamiltonian (\ref{ham2}) leads to
\begin{eqnarray}
& &{\cal H}=\sum_m\int d{\bf r} \Psi_m^\dagger({\bf r},t)
\left[\frac {{\bf p}^2}{2M}+U({\bf r})\right]\Psi_m({\bf r},t)
+\frac{\hbar}{2}\int d{\bf r}\{(c_0+c_2)
[\Psi_1^\dagger\Psi_1^\dagger\Psi_1\Psi_1
+\Psi_{-1}^\dagger\Psi_{-1}^\dagger\Psi_{-1}\Psi_{-1}
\nonumber\\
& &
+2\Psi_0^\dagger\Psi_0
(\Psi_1^\dagger\Psi_1+\Psi_{-1}^\dagger\Psi_{-1})]
+c_0\Psi_0^\dagger\Psi_0^\dagger\Psi_0\Psi_0
+2(c_0-c_2)\Psi_1^\dagger\Psi_1
\Psi_{-1}^\dagger\Psi_{-1}
+2c_2(\Psi_1^\dagger\Psi_{-1}^\dagger\Psi_0\Psi_0+H.c. )\}.
\label{ham22}
\end{eqnarray}

This form of the Hamiltonian is quite familiar in quantum optics, where it
describes four-wave mixing between a pump beam and two side-modes,
which are identified with the field operators $\Psi_0$ and $\Psi_{\pm 1}$
in the present situation. 

The three terms
in the two-body Hamiltonian which are quartic in one of the field operators
only, i.e. of the form $\Psi_i{^\dagger} \Psi_i^{\dagger} \Psi_i \Psi_i$,
can be readily interpreted as self-defocussing terms, corresponding to
the fact that the two-body potential is, for a positive scattering length
and a scalar field, analogous to a defocussing cubic nonlinearity in
optics. The terms involving two ``modes'', i.e. of the type
$\Psi_i^\dagger \Psi_i \Psi_j^\dagger \Psi_j$,
conserve the individual mode populations of the modes and simply lead to
phase shifts. Finally, the terms involving the central mode $\Psi_0$ and
{\em both} side-modes are the contributions of interest to us. They
correspond to a redistribution of atoms between the ``pump'' mode
$\Psi_0$ and the side-modes $\Psi_{\pm 1}$, e.g., by annihilating two atoms
in the central mode and creating one atom each in the side-modes. This is
the kind of interaction that leads to phase conjugation in quantum optics,
except that in that case the modes in question are modes of the
Maxwell field instead of the Schr\"odinger field. Note also that a similar
mechanism is at the origin of amplification in the Collective
Atom Recoil Laser (CARL) \cite{BonSal94,BonSalNar94,MooMey982}, see Sec.~V.

In the Hartree approximation, which is well justified for condensates
at $T=0$, the dynamics of a condensate described by the Hamiltonian 
(\ref{ham22}) is governed by the system of coupled nonlinear Schr\"odinger 
equations
\cite{LenMeyWri93,ZhaWalSan94,LenMeyWri94}
\begin{eqnarray}
& &i\dot{\phi}_{-1}({\bf r},t)=\frac{1}{\hbar}
\left[\frac{{\bf p}^2}{2M}+U({\bf r})
\right] \phi_{-1}+N\{c_2\phi_0\phi_0\phi_1^\star
+[(c_0+c_2)(|\phi_{-1}|^2+|\phi_0|^2)
+(c_0-c_2)|\phi_{1}|^2]\phi_{-1}\}
\nonumber\\
& &i\dot{\phi}_{0}({\bf r},t)=\frac{1}{\hbar}
\left[\frac{{\bf p}^2}{2M}+U({\bf r})
\right] \phi_0+N\{c_0|\phi_0|^2\phi_0
+(c_0+c_2)(|\phi_{-1}|^2+|\phi_1|^2)\phi_{0}
+2c_2\phi_1\phi_{-1}\phi_0^\star\}
\nonumber\\
& &i\dot{\phi}_{1}({\bf r},t)=\frac{1}{\hbar}
\left[\frac{{\bf p}^2}{2M}+U({\bf r})
\right] \phi_1+N\{c_2\phi_0\phi_0\phi_{-1}^\star
+[(c_0+c_2)(|\phi_{1}|^2+|\phi_0|^2)
+(c_0-c_2)|\phi_{-1}|^2]\phi_{1}\}.
\label{schro}
\end{eqnarray}
Consider for example a situation where the central mode, described by the
Hartree wave function $\phi_0$, is initially strongly populated while
the side-modes $\phi_{\pm 1}$ are weakly excited. In other words, we
consider the phase conjugation of a weak atomic beam from a large
condensate. It is then appropriate to introduce the matter-wave
optics equivalent of the undepleted pump approximation, whereby
\begin{equation}
{\dot {\phi}}_0 \simeq 0,
\label{undep}
\end{equation}
and the problem reduces to a set of coupled mode equations for
the two side-modes $\phi_{\pm 1}$, the central mode acting as a catalyst
for the coupling between them.

We take the trap potential $U({\bf r})$ to be of the harmonic form
\begin{equation}
U({\bf r})=M\omega_0^2 (x^2+y^2)/2
\end{equation}
that is, we assume that the dipole trap confines the atoms
in the transverse plane $(x,y)$, but not in the longitudinal direction
$z$. This geometry allows one to consider side-modes propagating along
that axis, rather than bouncing back and forth in an elongated trap.
In case of tight confinement in the transverse direction, we can assume
to a good approximation that the transverse structure of the condensate
is not significantly altered by many-body interactions and is
determined as the ground-state solution of the transverse potential.

Expressing the Hartree wave function associated with the hyperfine level
$m$ as
\begin{equation}
\phi_m({\bf r}, t) = \varphi_\perp (x,y) \varphi_m(z,t) e^{-i\omega_0 t},
\label{anzats0}
\end{equation}
we then have
\begin{equation}
\hbar\omega_0 \varphi_\perp(x,y)=\left [ -\frac{\hbar^2}{2M}
\left (\frac{\partial ^2}{\partial x^2}+\frac{\partial ^2}{\partial y^2}
\right)
+\frac {M\omega^2_0}{2}(x^2+y^2)\right]\varphi_\perp(x,y).
\end{equation}

The physical situation we have in mind is that of a weak ``probe'' in the
hyperfine state $m=-1$ propagating toward a large condensate in
state $m=0$ and at rest in the dipole trap, and generating a
backward-propagating conjugate matter wave in the hyperfine state $m = +1$.
Hence we express the longitudinal component of the Hartree
wave function as
$$
{\bbox{\varphi}}(z,t)\equiv\left(\begin{array}{c}
\varphi_{-1}(z,t) \\
\varphi_{0}(z,t) \\
\varphi_{1}(z,t)  \end{array}\right)
=\left(\begin{array}{c}\psi_{-1}(z,t) e^{-ikz} \\
2\psi_{0} \cos(kz) \\ \psi_{1}(z,t) e^{ikz}
\end{array}\right)e^{-i\omega t},
$$
\noindent where the slowly varying envelopes $\psi_m$ of the Hartree
wave function components $m = \pm 1$ satisfy the familiar inequalities
\begin{equation}
\left |\frac{\partial^2}{\partial z^2}\psi_m \right |\ll 
k\left |\frac{\partial}{\partial z}\psi_m \right |
\ll k^2 |\psi_m|.
\end{equation}
The ``pump'' wave function $\varphi_0$ is described
by a standing wave, a configuration that can be achieved
for instance by interfering two condensates \cite{AndTowMie97},
in a grating matter-wave
interferometer, \cite{OvcKozDen98} or in the CARL \cite{MooMey982}.
To first order in the probe and signal fields, this geometry
leads to a linearized system of two coupled-mode equations for the probe
and condensate fields. In the stationary state and projecting out
the transverse part of the wave function they reduce to
\begin{eqnarray}
i\frac{\hbar k}{2M}\frac{\partial}{\partial z}\psi_{-1}(z)&=&-N\eta
[2(c_0+c_2)\rho_0\psi_{-1}(z)
+c_2\psi_{0}^2 \psi_{1}^\star(z)] ,
\nonumber\\
i\frac{\hbar k}{2M}\frac{\partial}{\partial z}\psi_{1}^\star(z)&=&-N\eta
[2(c_0+c_2)\rho_0\psi_1^\star(z)
+c_2\psi_{0}^{\star2} \psi_{-1}(z)] ,
\label{phcon}
\end{eqnarray}
where  $\rho_0 = |\psi_0|^2$ and
\begin{equation}
\eta=\frac{\int dx dy |\varphi_\perp(x,y)|^4}{\int dx dy
|\varphi_\perp(x,y)|^2} .
\label{eff}
\end{equation}

These equations are of course familiar from optical
phase conjugation and their solution is well-known.
The evolution of the phase conjugate wave $\psi_1^\star$ contains
a term proportional to the density $\rho_0$ of the condensate  and the
field itself. In the absence of the second term, it would simply lead
to a phase shift of $\psi_1^\star$. Physically, it results from the
self-interaction of the conjugate field, catalyzed by the condensate
(pump) component. Its origin can be traced back to the
term proportional to $\Psi_1^\dagger\Psi_{1}^\dagger\Psi_0\Psi_0$ in
the Hamiltonian (\ref{ham22}).  The second term, in contrast, couples the
two side-modes via the condensate and is responsible for phase conjugation.

The general solution of Eqs. (\ref{phcon}) reads \cite{Fis83}
\begin{eqnarray}
\psi_{-1}(z)&=&\frac{e^{i\alpha z}}{\cos(|\kappa| L)}
\left(\right.- ie^{-i\beta}\sin(|\kappa|z)\psi_{1}^\star(L)
+\left.\cos(|\kappa|(z-L))\psi_{-1}(0) \right)
\nonumber\\
\psi_{1}(z) &=& \frac{e^{i\alpha z}}{\cos(|\kappa| L)}\left(
\right. \cos(|\kappa|z) \psi_1(L)
+i \left.e^{-i\beta}\sin(|\kappa|(z-L))\psi_{-1}^\star(0)\right),
\end{eqnarray}
where
$\alpha=2N\eta (c_0+c_2)\rho_0$,
$\kappa=\frac{N\eta c_2\psi_0^2}{\hbar k/2M}$ ,
and $e^{i\beta}=\kappa/|\kappa|$.

For the probe $\psi_{-1}(0)$ incident at $z=0$ and no incoming conjugate
signal $\psi_1(L)=0$, the conjugate wave in the input plane $z=0$ becomes
\begin{equation}
\psi_{1}(0)=-ie^{-i\beta}\tan(|\kappa|L)\psi_{-1}^\star(0),
\end{equation}
which demonstrates that the interaction of the probe and the
condensate results in the generation of a counterpropagating
phase-conjugated
signal.

In addition to its interest from a nonlinear atom optics point of view,
matter-wave phase conjugation could also be used as a diagnostic tool
for Bose-Einstein condensates. For instance, we noted that the parameter
$|\kappa|L$ is proportional to the difference in scattering lengths between
the singlet and triplet states. Hence, this quantity could in principle
be inferred from phase conjugation measurements.

\section{The binary collision atom laser}

As a second example illustrating the close analogy between nonlinear optics
and nonlinear atom optics, we now consider the atom laser. There has been
quite a bit of confusion about what is meant by an ``atom laser'' in the
past, apparently associated with the fact that since the number of atoms is
conserved, one cannot possibly achieve their amplification.\footnote{Another
problem is that the L in the acronym for laser stands for
Light. But this is history repeating itself: in the early days of lasers,
they were commonly called ``optical masers.''}
But of course, this is not the point. The main purpose of an atom laser
is to create a coherent, macroscopic population in a given center-of-mass
mode of atomic motion, such as to create a coherent atomic beam. Clearly,
the number of particles in a given mode needs not be conserved: it can readily
be amplified by cleverly extracting atoms from a large reservoir.

A primitive atom laser has been demonstrated experimentally by the
MIT group, which outcoupled a fraction of a sodium condensate from a
magnetic trap via rf coupling between the weak-field-seeking $m_F=-1$
state and the strong-field-seeking $m_F = 1$ state
\cite{MewAndKur97,AndTowMie97}. We concentrate here
on a different scheme, which has so far not been realized experimentally,
the binary collision atom laser
\cite{GuzMooMey96,WisMarWal96,HolBurGar96,MooMey972,ZobMey98}.
In this system,
the matter-wave resonator consists, e.g., of an optical dipole trap.
In order to concentrate on the essential dynamics only three out of the
multitude of trap modes are taken into account explicitly.

The atom laser then operates as follows: Bosonic atoms
in their ground electronic state are incoherently pumped into a trap
level of ``intermediary'' energy (mode 1). There they undergo binary
collisions which take one of the atoms involved to the tightly bound
laser mode 0, whereas the other one is transferred to the heavily damped
loss mode 2. This latter atom leaves the resonator quickly, thereby
providing the irreversibility of the pumping process. A macroscopic
population of the laser mode can build up as soon as the influx of
atoms due to pumping compensates for the losses induced by the damping.

In the description of this laser scheme one has to take into
account that in addition to the pumping collisions other types of
interatomic collisions can also occur. These considerations lead to an
ansatz for the atom laser master equation of the form
\begin{equation} \label{al3m}
\dot{W} =  -\frac{i}{\hbar}
[H_0+H_{col},W] +\kappa_0{\cal D}[c_0]W+\kappa_1(N+1){\cal D}
[c_1]W +\kappa_1 N {\cal D}[c_1^{\dagger}]W + \kappa_2{\cal D}[c_2]W .
\end{equation}
In this equation, we use the second quantized formalism in which each
center-of-mass atomic mode is associated with a bosonic annihilation operator
$c_i$, and $W$ denotes the atomic density operator.\footnote
{Since we consider ground state atoms only, they
are fully described by their center-of-mass quantum numbers.}
The free Hamiltonian is given by
$$H_0=\sum_{i=0,1,2}\hbar \omega_i c_i^{\dagger} c_i,$$
$\omega_i$ being the mode frequencies. The general form of the collision
Hamiltonian can be written as
\begin{equation}\label{hcol}
H_{col}=\sum_{i\leq j,k\leq l} \hbar V_{ijkl} c^{\dagger}_i c^{\dagger}_j
c_k c_l
\end{equation}
with $V_{ijkl}$ the matrix elements of the two-body interaction
Hamiltonian responsible for the collisions. It is sufficient for the present
purpose to restrict our attention to the reduced form
\begin{equation}
\label{hcolres}
H_{col}  =  \hbar(V_{0211}c^{\dagger}_0 c^{\dagger}_2 c_1 c_1 + V_{1102}
c^{\dagger}_1 c^{\dagger}_1 c_0 c_2 + V_{0000} c^{\dagger}_0
c^{\dagger}_0 c_0 c_0 + V_{0101} c^{\dagger}_0 c^{\dagger}_1 c_0 c_1 +
V_{1111} c^{\dagger}_1 c^{\dagger}_1 c_1 c_1)
\end{equation}
in which (besides the pumping collisions) only those collisions are
retained which are expected to have the most significant influence on
the phase dynamics.

It is worth emphasizing that there is a fundamental difference
in the way collisions are handled in conventional atomic physics and in the
present situation. Usually, and for instance in the phase conjugation
problem of the preceding section, collisions are handled in terms of
scattering amplitudes between initial and final states. In matter-wave
resonators, however, such an approach becomes meaningless. Rather, the
collisions are seen to induce transitions between cavity modes. Hence, the
collision Hamiltonian (\ref{hcolres}) takes the form of a {\em mode-coupling}
Hamiltonian between three cavity modes. It is therefore mathematically
closely related to that of Eq.~(\ref{ham22}), except that the coupling is
now between center-of-mass modes, rather than magnetic sublevels.
Another difference is of course that in phase conjugation, we are interested
in a coherent interaction, and neglect the effects of dissipation, which are
known to be detrimental to the achievement of phase reversal. In contrast,
the laser and the atom laser are open systems that rely explicitly on the
presence of pump and probe mechanisms to achieve steady-state operation.

The damping rates of the
cavity modes are given by the coefficients $\kappa_i$, and the strength of
the external pumping of mode 1 is characterized by the parameter $N$,
which is the mean number of atoms to which mode 1 would equilibrate in
the absence of collisions. The superoperator ${\cal D}$ is defined by
\begin{equation}
{\cal D}[a] P = aPa^{\dagger}-\textstyle{\frac 1 2}(a^{\dagger} a P +
P a^{\dagger} a)
\end{equation}
with arbitrary operators $a$ and $P$.

In order to achieve a sufficiently high degree of irreversibility it is
necessary that $\kappa_2$ is much larger than the damping rates of the
other modes. This suggests adiabatically eliminating
this mode, an approximation that leads to the simplified master equation
\cite{GuzMooMey96,WisMarWal96,HolBurGar96}
\begin{equation} \label{al2m}
\dot{\rho}  =  -\frac{i}{\hbar}[H_{c},\rho]+\kappa_0{\cal D}[c_0]\rho
+\kappa_1(N+1){\cal D} [c_1]\rho \nonumber \\
+\kappa_1 N {\cal D}[c_1^{\dagger}]\rho + \Gamma {\cal D}
[c_0^{\dagger}c_1^2]\rho.
\end{equation}
Equation (\ref{al2m}) is written in the interaction picture with respect
to $H_0=\hbar\omega_0 c^{\dagger}_0 c_0 + \hbar\omega_1 c^{\dagger}_1 c_1$,
and the reduced density matrix $\rho$ is
$\rho=\mbox{Tr}_{\mbox{{\small mode 2}}}[W]$.
The reduced collision Hamiltonian $H_c$ is
\begin{equation}
H_c = \hbar(V_{0000}c_0^\dagger c_0^\dagger c_0 c_0 +
V_{1111}c_1^\dagger c_1^\dagger c_1 c_1) ,
\end{equation}
and $\Gamma=4|V_{0211}|^2/\kappa$. Consistently
with Ref.~\cite{WisMarWal96} we call the limiting cases
$\Gamma \ll \kappa_0$ and $\Gamma \gg \kappa_0$ the weak and strong
collision regimes, respectively. The reduced master equation (\ref{al2m})
forms the basis of most studies of binary collision atom lasers.

The master equations (\ref{al3m}) and (\ref{al2m}) can easily be solved
numerically using standard quantum Monte Carlo simulation techniques
\cite{DumParZolGar92,MolCasDal93}. This is discussed in detail
in Ref.~\cite{MooMey972}.
Figure 1 shows the probability $P(n)$ of having $n$ atoms in the laser mode
as a function of time for the case where only $V_{0211}=V^\ast_{1102}$
is taken to be non-zero. The parameters used are $\kappa_0=0.01\kappa_1$,
$N=2$, $\kappa_2=10\kappa_1$ and $V_{0211}=2\kappa_1$,
the time is in units of $\kappa_1^{-1}$.
We see that the population of the
lasing mode builds up from zero to a state that suggests a Poissonian
distribution with a mean number of atoms $n_0=40$. In contrast, the evolution
of the pump and decay modes (not shown) shows no significant build-up of
population in steady state.

Among the most important characteristics of a laser are the coherence
properties of its output, in particular its linewidth.
In order to obtain an analytical approximation for the laser linewidth
in the two-mode system (\ref{al2m})
a linearized fluctuation analysis can be performed
\cite{MeySar91,WalMil94}.
Assuming as usual that the first-order correlation function $C_0(\tau)=
\langle c_0^{\dagger}(\tau) c_0(0) \rangle$ is determined only by the
phase fluctuations one obtains in this way
\begin{equation}\label{defc0}
C_0(\tau)= \bar{n}_0 e^{-i \bar{\phi}_0} \exp[-\textstyle{\frac 1 2}
\sigma_{\phi_0}(\tau)] \label{c0},
\end{equation}
where the equilibrium populations of laser and pump mode are given by
\begin{equation}
{\bar n}_0 = \frac{1}{2} \frac{\kappa_1}{\kappa_0} (N - {\bar n}_1) ,\
\end{equation}
\begin{equation}
{\bar n}_1 = \sqrt{\kappa_0/\Gamma} ,
\end{equation}
the threshold condition being $N > \sqrt{\kappa_0/\Gamma}$.
The time-dependent deterministic phase drift is indicated by $\bar{\phi}_0$.
This phase drift leads to a shift of the center of the power
spectrum (the Fourier transform of the correlation function) by an
amount $2V_{0000}\bar{n}_0+V_{0101}\bar{n}_1$ with respect to the
collisionless case. The behavior of the correlation function $C_0(\tau)$
is thus essentially determined by the coefficient
$\sigma_{\phi_0}(\tau)$ which is the covariance of the phase
fluctuations in the laser mode.

This coefficient can be evaluated explicitly in a straightforward way, 
however the ensuing expression is rather complicated. In the following we 
restrict the discussion to two limiting cases which illustrate the essential 
aspects of the influence of the elastic collisions on the laser linewidth.
(It should also be noted at this point that
$C_0(\tau)$ does not depend on elastic collisions between pumping mode
atoms, which are characterized by the parameter $V_{1111}$.)

({\it a}) $V_{0101}=0$.
Equation (\ref{defc0}) implies that the most important aspects of the
correlation function can be inferred from the study of $\sigma_{\phi_0}(\tau)$
in the time interval where it is smaller than or of the order of unity.
Examining the behavior of $\sigma_{\phi_0}(\tau)$ as a function of
$V_{0000}$ (with all other parameters kept constant) one can distinguish
between two different regimes. For small $V_{0000}$
the time evolution of $\sigma_{\phi_0}(\tau)$ relevant for $C_0(\tau)$
is well approximated by
\begin{equation}\label{g00small}
\sigma_{\phi_0}(\tau)\simeq \tau [w + \kappa_0/(2\bar{n}_0)],
\end{equation}
where 
\begin{equation}
w=V_{0000}^2\frac{\kappa_1 N}{\kappa_0^2} \left(2+2\sqrt{\kappa_0/
\Gamma}+\frac{N}{N-\sqrt{\kappa_0/\Gamma}}\right).
\end{equation}
For such values of $V_{0000}$, $C_0(\tau)$ decays therefore
exponentially, and the power spectrum is Lorentzian. If $w \gg \kappa_0/
(2\bar{n}_0)$ the linewidth is proportional to $V_{0000}^2\bar{n}_0$. In
contrast to the situation with conventional lasers, it {\em increases}
linearly with the number of atoms in the laser mode.

In case $V_{0000}>\sqrt{\frac{8}{\ln 2}}\frac{\sqrt{\bar{n}_0}}{w'}$
with $w'=w/V_{0000}^2$
we find a different behavior of the correlation function.
Under these circumstances it decays like a Gaussian, i.e.,
\begin{equation}\label{g00large}
\sigma_{\phi_0}(\tau)\simeq4V_{0000}^2\bar{n}_0\tau^2.
\end{equation}
The spectrum is thus itself also of Gaussian shape and its linewidth
proportional
to $V_{0000}\sqrt{\bar{n}_0}$. The atom laser linewidth still increases
with ${\bar n}_0$, albeit less dramatically than in the preceding case.

({\it b}) $V_{0000}=0$. In this case the expansion of $\sigma_{\phi_0}(\tau)$
to leading order in $\bar{n}_0$ yields
\begin{equation}\label{g01app}
\sigma_{\phi_0}(\tau)=\left(\frac{V_{0101}^2}{2\Gamma \bar{n}_0}+
\frac{\kappa_0}{2\bar{n}_0}\right) \tau
\end{equation}
The correlation function thus decays exponentially for all values of
$V_{0101}$. A qualitative change in behavior as in the previous
situation does not occur. The linewidth of the spectrum is now proportional
to $V_{0101}^2/\bar{n}_0$, and becomes {\em narrower} when the
population of the laser mode is increased, very much like
the familiar Shawlow-Townes linewidth of conventional lasers.
The analytical estimates of Eqs.~(\ref{g00small}), (\ref{g00large}), and
(\ref{g01app}) are in good agreement with numerical
quantum Monte Carlo simulations \cite{ZobMey98}.

\section{Parametric amplification: the collective atom recoil laser}

As a final example of nonlinear atom optics in ultracold atomic systems,
we briefly review some of the most salient aspects of the ultracold atoms
operation of the Collective Atomic Recoil Laser, or CARL \cite{BonSal94}.
This device
consists of three main components: (1) the active medium, which
consists of a gas of two-level atoms, (2) a strong pump
laser which drives the two-level atomic transition, and (3)
a ring cavity which supports an electromagnetic mode (the probe)
counterpropagating with respect to the pump.
What makes the CARL interesting is that the initial state, consisting of
a thermal cloud of atoms and no photons in the cavity,
is exponentially unstable.
Laser oscillations appear spontaneously in the probe mode correlated
with the appearance of a density modulation in the atomic sample.
The original purpose of
the CARL, which operates then at room or higher temperatures, was the
generation of a tunable coherent light field from atoms in a way similar to
light amplification in free-electron lasers, i.e., gain correlated with
bunching and in the absence of population inversion.
But as we shall see, when operating
at ultracold temperatures it can simultaneously parametrically amplify
spatial side-modes of a Bose-Einstein condensate. This is the aspect of
the CARL that we concentrate on in this section.

In the absence of collisions, the second-quantized Hamiltonian of a sample
of two-level atoms interacting with a classical pump laser
and a counterpropogating probe cavity mode is
\begin{equation}
{H}=\sum_k {H}(k)+\hbar cq{A}^\dag{A}.
\label{hatH}
\end{equation}
Here ${H}(k)$ is given by
\begin{eqnarray}
{H}(k)&=&\frac{\hbar^2 k^2}{2m}{c}^\dag_g(k){c}_g(k)
+\left(\frac{\hbar^2k^2}{2m}+\hbar\omega_0\right)
{c}^\dag_e(k){c}_e(k) \nonumber\\
&+&\left[\hbar\frac{\Omega^\ast_L}{2}e^{i\omega_Lt}{c}^\dag_g(k+k_L){c}_e(k)
+ i\hbar g{A}^\dag{c}^\dag_g(k-q){c}_e(k) + H.c.\right],
\label{Hk}
\end{eqnarray}
where the bosonic matter wave  operator ${c}_g(k)$ annihilates
a ground state atom of momentum $\hbar k$, and ${c}_e(k)$
annihilates an excited atom of momentum $\hbar k$, with
\begin{equation}
[{c}_g(k),{c}^\dag_g(k^\prime)]=
[{c}_e(k),{c}^\dag_e(k^\prime)]=\delta_{kk^\prime},
\label{comm}
\end{equation}
all other matter-wave commutators being equal to zero.
The pump is characterized by its frequency $\omega_L$, its
Rabi frequency $\Omega_L$, and its wavenumber $k_L$.
The probe field operator $A$ annihilates a photon with wavenumber $q$,
and the atom-probe coupling constant is $g=d[cq/(2\epsilon_0LS)]^{1/2}$,
where $d$ is the atomic dipole moment, and $LS$ is the cavity volume.
Note that the atomic recoil
is explicitly included in the electric dipole interaction coupling the
electronic ground and excited states.

The Hamiltonian (\ref{Hk}) readily yields
the Heisenberg equation of motion for the probe field mode
as
\begin{equation}
\frac{d}{dt}{A}=-icq{A}+g\sum_k{c}^\dag_g(k-q){c}_e(k).
\label{maxwell2}
\end{equation}
Hence, all that is required to determine the field evolution are
bilinear combinations of atomic creation and annihilation operators.
Their equations of motion are in turn
\begin{equation}
\frac{d}{dt} {c}^\dag_g(k){c}_e(k^\prime) =
\frac{i}{\hbar} [{ H}, {c}^\dag_g(k){c}_e(k^\prime)].
\label{emo}
\end{equation}

Because the Hamiltonian (\ref{Hk}) is bilinear in the atomic operators,
we observe readily that the equations of motion (\ref{emo}) involve only
bilinear combinations. This is a direct consequence of the fact that
collisions are neglected in this model. In the language of manybody theory,
this means that the BBGKY hierarchy is exactly truncated. If the
probe field
is treated classically, then Eq. (\ref{maxwell2}) becomes a c-number
equation involving only the matter-wave expectation values
$\langle{c}^\dag_g(k-q){c}_e(k)\rangle$, which are nothing but
the single-particle atomic density matrix elements. Since
these elements are coupled only to other single-particle density matrix
elements, the system can be solved without any truncation scheme being
required, albeit numerically in general. If instead of treating
probe field classically we had chosen to treat the atomic single-particle
density matrix classically, we would have reached the same result.
Thus converting Eqs. (\ref{maxwell2}) and (\ref{emo}) into c-number equations
is equivalent to a classical field theory for both the atomic and optical
fields. This approach  is discussed in detail
in Ref. \cite{MooMey982}, and we do not pursue it further here.

Since we are interested in the interaction between
light and a Bose condensate, we must be careful that electromagnetic heating
does not occur. Hence, the optical fields must be far off-resonant
from any electronic transition, and the upper electronic levels can be
adiabatically eliminated. The atoms are then described as a scalar field,
since only their electronic ground state remains. In that case
one finds the system is described by the effective Hamiltonian
\begin{eqnarray}
{H}&=&\frac{\hbar^2}{2m}\sum_kk^2{c}^\dag_g(k){c}_g(k)+\hbar cq{A}^\dag{A}
+i\frac{\hbar}{2\Delta_L}\sum_k\left[g\Omega_Le^{-i\omega_Lt}
{A}^\dag{c}^\dag_g(k-2k_0){c}_g(k) - H.c.\right]
\nonumber \\
&+&\frac{\hbar}{\Delta_L}\left(\frac{|\Omega_L|^2}{4}
+|g|^2{A}^\dag{A}\right)
\sum_k{c}^\dag_g(k){c}_g(k),
\label{HeffCARL}
\end{eqnarray}
where we have introduced the recoil kick $2k_0=q+k_L$, and the pump
detuning $\Delta_L=\omega_L-\omega_a$.

If the atomic sample is initially a condensate at zero temperature, the
dominant mode, at least for short times, is the $\kappa = 0$ mode. It is
macroscopically populated, while all other matter wave modes are in a
vacuum. The equation of motion for the condensate mode is
\begin{equation}
\frac{d}{dt}{c}_g(0)=\frac{g}{2\Delta_L}\left[
\Omega^\ast_Le^{-i\omega_Lt}{A}^\dag{c}_g(2k_0)
-\Omega_Le^{i\omega_Lt}{A}{c}_g(-2k_0)\right]
\end{equation}
This equation indicates that the condensate mode is coupled via the
electromagnetic field to the two side-modes  at $\pm 2k_0$.
In addition, each side mode is coupled to its neighbors
at intervals of $\pm 2k_0$. This is
mathematically similar to the situation of Sec.~III, see Eq. (\ref{schro}),
with, however, three major differences: (a) in the case of phase conjugation,
the mode coupling is between different magnetic sublevels of the atomic
field, while it is now between momentum states; (b) the coupling is now
due to a Raman-like coupling induced by two counterpropagating fields,
instead of ground-state collisions. Hence the mode coupling
now involves an infinite manifold of matter-wave modes and two
electromagnetic fields modes ---
one of them treated classically in the present example --- instead of
four matter-wave modes. This is of course a direct consequence of the fact
that collisions are the result of a phenomenological approach, eliminating
the electromagnetic vacuum as we have seen in Sec.~II. (c) Instead
of three magnetic sublevels, we now have an infinite hierarchy of
center-of-mass modes of the matter-wave  field to deal with.

This system is similarly related to the atom laser of Sec.~IV, except that
in that case, matter-wave modes are selected by the atomic resonator, as
well as possibly by the selection rules of the collision Hamiltonian
\cite{MooMey971}, whereas here they are selected from the continuum of
modes simply by conservation of momentum. In
addition, there are no atomic pump and decay mechanisms in the present
model, which describes a fully coherent interaction.

Let us now discuss this hierarchy in some more detail. We note that each
mode $k$ is directly coupled only to its neighboring modes
$k \pm 2k_0$. But except for the condensate mode $k = 0$, all
modes are initially empty, so in the early stages the dominant dynamics
results from the coupling between the condensate mode and its two neighboring
modes. Neglecting then the higher-order modes, and further treating the
condensate mode as a constant c-number, a sort of undepleted pump approximation
for a classical atom-laser field and an excellent approximation at $T=0$
and for a sufficiently large condensate, we find that we have reduced the
system to a linear three-mode
problem. It is easily shown that this reduced problem can be described
by the effective Hamiltonian
\begin{equation}
{H}=4\hbar\omega_R\left({c}^\dag_-{c}_-+{c}^\dag_+{c}_+
-\Delta{a}^\dag{a}
+\chi\left[{a}^\dag{c}^\dag_-+{a}^\dag{c}_+
+{c}^\dag_+{a}+{c}_-{a}\right]\right),
\label{heff}
\end{equation}
where $\omega_R=\hbar k_0^2/2m$ is the atomic recoil frequency,
\begin{equation}
{c}_\pm=e^{i|\Omega_L|^2t/4\Delta_L}{c}_g(\pm 2k_0),
\label{defcpm}
\end{equation}
\begin{equation}
{a}=-i(g^\ast\Omega^\ast_L\Delta_L/|g||\Omega_L||\Delta_L|)
e^{i\omega_Lt}{A},
\label{defa}
\end{equation}
\begin{equation}
\chi=\frac{|g||\Omega_L|}{8\omega_R\Delta_L}\sqrt{N},
\label{defalpha}
\end{equation}
\begin{equation}
\Delta=\frac{\omega_L-\omega}{4\omega_R},
\label{defDelta}
\end{equation}
$\omega=cq-|g|^2N/\Delta_L$ and $N$
is the mean number of atoms in the condensate.
We see that $\chi^2$ is an intensity parameter, proportional to
the product of the intensities of the pump laser and the
initial condensate, and $\Delta$ is simply the pump-probe detuning in
units of $4\omega_R$.
The Hamiltonian (\ref{heff}) gives the full quantum field theory
description of the zero temperature CARL, and is valid for all times
short enough so that $\langle{c}^\dag_\pm{c}_\pm\rangle\ll N$
and $\langle{a}^\dag{a}\rangle \ll |\Omega_L|^2/4|g|^2$.

The presence of terms such as
${a}^\dag{c}^\dag_-$ in Eq. (\ref{heff})
immediately brings to mind the non-degenerate optical
parametric amplifier \cite{MeyWal91} which generates
highly non-classical optical fields. These fields exhibit two-mode
intensity correlations and squeezing, and have been extensively employed
in the creation of EPR photon pairs for fundamental studies of the laws of
quantum mechanics. Our system represents a generalization
in that we now have three entangled quantum fields, and is especially
interesting in that two of the fields are atomic rather than optical.
A detailed analysis of quantum mode coupling in the CARL can be found in
Ref. \cite{MooMey983}. Here we focus on the intensity
dynamics and fluctuations.

The dynamics of the system can be determined by solving the three coupled
mode equations
\begin{equation}
\frac{d}{d\tau}
\left(\begin{array}{c}
{a}(\tau)\\ {c}^\dag_-(\tau)\\ {c}_+(\tau)\\ \end{array}\right)
=i\left(
\begin{array}{ccc}
-\Delta&-\chi&-\chi\\
\chi&1&0\\
-\chi&0&-1\\
\end{array}\right)
\left(\begin{array}{c}
{a}(\tau)\\ {c}^\dag_-(\tau)\\ {c}_+(\tau)\\ \end{array}\right),
\label{eqs}
\end{equation}
where we have introduced the dimensionless time variable $\tau=4\omega_R t$.

The spectrum of (\ref{eqs}) has been studied in detail in
\cite{MooMey981,MooMey982}, with the result that under certain threshold
conditions the eigenvalues take the form $\lambda_1=\omega_1$,
$\lambda_2=\Omega+i\Gamma$, and $\lambda_3=\Omega-i\Gamma$,
where $\omega_1$, $\Omega$, and $\Gamma$ are all real quantities.
This means that the time-dependance of the operators will, after some
transients, grow exponentially in time at the CARL growth rate $\Gamma$.

If we now consider an initial state where the atomic side modes are in the
vacuum state and the probe mode has been injected with an initial
field in a coherent state $|\alpha\rangle$,
we find that after the transients have died away
the mean intensities of the three modes are given by
\begin{equation}
I_A(\tau)\equiv\langle{a}^\dag(\tau){a}(\tau)\rangle
\approx(|\alpha|^2R_{11}^2+R_{12}^2)e^{2\Gamma\tau},
\label{IA}
\end{equation}
\begin{equation}
I_-(\tau)\equiv\langle{c}^\dag_-(\tau){c}_-(\tau)\rangle
\approx(|\alpha|^2R_{21}^2+R_{22}^2)e^{2\Gamma\tau},
\label{Iminus}
\end{equation}
and
\begin{equation}
I_+(\tau)\equiv\langle{c}^\dag_+(\tau){c}_+(\tau)\rangle
\approx(|\alpha|^2R_{31}^2+R_{32}^2)e^{2\Gamma\tau}.
\label{I+}
\end{equation}
Here the coefficients $\{R_{ij}\}$ are given by
$R_{ij}=|v_{i3}v^{-1}_{3j}|$, where the columns of the matrix $v$ are the
eigenvectors of the $3\times 3$ linear system described by Eq.\ (\ref{eqs}).
For a given set of control parameters $\chi$ and $\Delta$ these coefficients
are simply constants, whose analytic expressions, while straightforward
to derive, are too unwieldy to reproduce here.
Thus we see that the intensities have a stimulated component, proportional
to $|\alpha|^2$,
and a spontaneous component, which is present even when all three
modes begin in the vacuum state. In this case the system is triggered by
quantum noise in the form of fluctuations in the atomic bunching.

The second-order equal-time intensity correlation function is defined,
e.g. in the case of the probe field, as
\begin{equation}
g^{[2]}(\tau)=\frac{
\langle{a}^\dag(\tau){a}^\dag(\tau){a}(\tau){a}(\tau)\rangle}
{\langle{a}^\dag(\tau){a}(\tau)\rangle^2}.
\label{defg2}
\end{equation}
This gives a measure of the intensity fluctuations, and hence the
coherence properties of the various modes.
After the transient regime, the correlation functions for all three
modes are given by
\begin{equation}
g^{[2]}(\tau)\approx 2-\frac{|\alpha|^4}{[|\alpha|^2+f(\chi,\Delta)]^2},
\label{g2}
\end{equation}
where the fluctuation function $f(\chi,\Delta)=
|v^{-1}_{32}/v^{-1}_{31}|^2$
is approximately unity near the region on the $\Delta$ axis where
the exponential growth rate $\Gamma$ is maximized for fixed $\chi$,
and steadily increases as one moves away from this region.
We see that in the spontaneous case $(|\alpha|^2=0)$ 
the intensity correlation
functions are equal to 2, which is characteristic of a thermal, or chaotic
field. However, as $|\alpha|^2$ is increased, $g^{[2]}$ approaches 1,
which signifies a coherent, or Poissonian field. Thus we see that by varying
readily adjustable experimental parameters, such as the injected probe
field strength
and frequency, the condensate atom number, and/or the pump intensity and
detuning, we have the capability to vary the intensity fluctuations of the
generated fields continuously between thermal and coherent limits.

\section{Outlook}

In summary, then, we see that atom optics is progressing along a path that
closely parallels that of optics following the invention of the
laser. In many situations, it is possible to understand an atom optical effect
by simply reversing the roles of light and matter, and indeed, much of the
inspiration leading to atom lasers and nonlinear atom optics results from
such an approach. The pioneering work of Peter Franken and his
colleagues
play a central role in these developments. But what would probably intrigue
Peter most is not so much the way in which atom optics and optics are 
similar, but rather those important aspects in which they differ, as results in
particular from the fact that atoms have an internal structure, are
massive, and are composite particles. Work along these lines will no doubt
keep him amused for many years.

\acknowledgements
This work is supported in part by the U.S.\ Office of Naval Research
Contract No.\ 14-91-J1205, by the National Science Foundation Grant
PHY95-07639, by the U.S.\ Army Research Office and by the
Joint Services Optics Program.


\begin{figure}
\caption{Probability $P(n_0)$ for having $n_0$ atoms in mode $0$ as a function
of time. The parameters for this plot are
$\kappa_0=.01$, $\kappa_1=1$, $N=2$, $\kappa{2}=10$, $V_{0211}=2$, and
$V_{0000}=V_{0101}=V_{1111}=0$, all given in units of $\kappa^{-1}_1$.}
\end{figure}

\end{document}